\documentclass[a4paper,10pt, twoside]{article}
\usepackage{a4wide}
\usepackage{amssymb}
\usepackage{amsmath}
\usepackage{amsthm}
\usepackage{graphicx}
\usepackage{fancyhdr}
\usepackage{datetime}

\title{Relaxation dynamics in the gapped XXZ spin-1/2 chain}
\author{Jorn Mossel\footnote{J.J.Mossel@uva.nl} and Jean-S\'ebastien Caux\footnote{J.S.Caux@uva.nl}\\Institute for Theoretical Physics, Universiteit van Amsterdam
\\Valckenierstraat 65, 1018 XE Amsterdam, The Netherlands. }
\date{\ddmmyyyydate \today}

\begin{document}
\maketitle
\begin{abstract}
\noindent
We study the dynamics of a quench-prepared domain wall state released into a system whose unitary time evolution is dictated
by the Hamiltonian of the Heisenberg spin-$1/2$ gapped antiferromagnetic chain.  Using exact wavefunctions and their overlaps with
the domain wall state allows us to describe the release dynamics to high accuracy, up to the long-time limit, for finite as
well as infinite systems.  
The results for the infinite system allow us to rigorously prove that the system in the gapped regime ($\Delta >1$) cannot thermalize
in the strict sense. 
\end{abstract}

\section{Introduction}

Statistical mechanics, be it for classical or quantum many-body systems, relies on the reasonable assumption that all microstates of a system
in equilibrium are equally likely to be realized.  In order to be correctly described by a statistical ensemble, a system must therefore be
either coupled to an infinite thermal bath, or (if isolated) effectively display some form of 
ergodicity as per the eigenstate thermalization hypothesis \cite{Deutsch1991,Srednicki1994}, 
thereby allowing it to forget its initial state by `relaxing' towards a well-defined equilibrium state independent of the starting 
conditions.  The existence of these relaxation processes is not usually questioned, but their precise mechanisms occur in a generally unspecified manner.
For individual systems, classical mechanics distinguishes integrable and non-integrable cases,
the former possibly having stable quasiperiodic orbits protected by the KAM theorem \cite{KAM}, and therefore showing no true ergodicity.  For 
non-integrable cases, ergodicity is taken for granted as a consequence of chaotic dynamics.

In the quantum case, the question of equilibration has been the subject of many recent 
works, focusing in particular on the time evolution of a system after a quantum quench, at which a global parameter 
in the Hamiltonian is suddenly changed \cite{Calabrese2006} - \cite{Biroli2009}.  This is equivalent to releasing a prepared state and letting it evolve 
unitarily in time according to the Hamiltonian after the quench.  Such situations, some of which are experimentally realizable, lead
to many interesting theoretical questions.  
Under what circumstances is there a well-defined state at large times ({\it i.e.} what, if anything, remains of the initial conditions for given quench situations)?  
Can a general theory of relaxation based purely on the details of quantum dephasing be formulated?
Are there different classes of systems ({\it e.g.} integrable versus non-integrable, finite versus infinite, gapped versus gapless) having different generic behaviour?  
What remains of the usefulness of commonly used quasiparticle bases to understand the dynamics after the quench?  
Clearly, much progress remains to be done in order to have a full understanding of
all these issues.

It is our purpose here to consider a well-defined situation, which can be handled to a sufficient degree of exactness to provide some partial
but reliable insights into some of these questions.  The approach we will consider is based on the use of exact wavefunctions for the 
Heisenberg magnet, and therefore exploits the integrability of this system.  On the one hand, this renders some of our results non-generic.
On the other hand, we are able to provide hard facts for finite as well as infinite systems, gapped versus gapless, and exclude some possible scenarios.
Since the theoretical understanding of the nonequilibrium dynamics of strongly-correlated systems is still in its infancy, such
example cases will hopefully provide worthwhile reference points for later developments.

The paper is organized as follows.  In section 2, we first define our nonequilibrium problem, and present the tools we will use to study it.
We then discuss in section 3 the set of eigenstates we use to obtain quantitative results.  Section 4 considers the work probability distribution resulting 
from the quench for various system parameters.  In section 5, we study relaxation dynamics by concentrating on the Loschmidt echo, which we compute.  After a few words on the thermodynamic limit, we offer a discussion of our results and present our conclusions.  
All technical details for the computations are relegated to a series of appendices, which can safely be skipped by 
the reader only interested in the final results and conclusions.  On the other
hand, these appendices explain all calculations in sufficient detail to be reproduced by the specialist reader.

\section{Formulation}
We consider an isolated spin chain of $N$ sites, each occupied by a local spin-$1/2$ degree of freedom.  For definiteness, we put the spins on a ring
and impose periodic boundary conditions.  The thought experiment we perform consists in preparing the quantum state of the system at $t = 0$ as
\begin{equation}
|\phi\rangle =  |\underbrace{\downarrow \ldots \downarrow}_{M} \underbrace{\uparrow \ldots \uparrow}_{N-M} \rangle.
\label{domainwallstate}
\end{equation}
This state thus contains a magnetic domain wall between sites $M$ and $M+1$, and another (anti-) domain wall between sites $N$ and $1$ (in view
of the periodic boundary conditions).  This state can be prepared in different ways:  we can view it as being created by an Ising model with an appropriate 
position-dependent field, or as resulting from a sudden polarizing pulse applied on a section of an initially fully polarized chain.
For times $t > 0$, we let this state evolve unitarily in time under the antiferromagnetic XXZ Hamiltonian
\begin{equation}
H_{XXZ} = J \sum_{j=1}^N \left[ \frac{1}{2\Delta} \left(S_{j}^-S_{j+1}^+  + S_{j}^+S_{j+1}^- \right) + S_{j}^z S_{j+1}^z \right].
\end{equation}
Note that the Hamiltonian is rescaled by a factor of $1/\Delta$ as compared to how it usually appears in the literature, in order the have 
a well-defined Ising limit ($\Delta\rightarrow\infty$).
For $\Delta >1$ the spectrum of this theory is gapped, while for $-1 < \Delta \leq 1$ the system is in the quantum critical regime.  We consider the case when the system is antiferromagnetic (i.e. $J>0$). In order to simplify the expressions and without loss of generality we put $J=1$ throughout the paper.

The initial domain wall state (\ref{domainwallstate}) is not an eigenstate of the XXZ Hamiltonian away from the Ising limit $\Delta \rightarrow \infty$.
On the other hand, for any fixed value of $\Delta$, the exact eigenstates of this model form a basis in the Hilbert space on which we can 
at least in principle decompose the
initial domain-wall state to arbitrary accuracy.  Given this decomposition, the solution of the Schr\"odinger equation
becomes straightforward, and we obtain the exact time-dependent wavefunction after the quench as the linear decomposition
\begin{equation}\label{timeevolphi}
|\phi (t)\rangle = \sum_{n}  e^{-iE_n t}Q_n |\Psi_n\rangle, \hspace{3cm}  Q_n \equiv \langle \Psi_n|\phi\rangle,
\end{equation}
where the sum is over all the $\left( \begin{array}{c} N \\ M \end{array} \right)$ eigenstates of $H_{XXZ}$ at fixed total magnetization, 
and the complex amplitudes $Q_n$ represent the overlaps (vector of the quench matrix) between the starting state and the exact eigenfunctions. 
Since we always work with normalized wavefunctions, the coefficients $Q_n$ should by definition satisfy the constraint 
\begin{equation}
\sum_{n} |Q_n|^2 =1.
\label{sumrule}
\end{equation}
This constraint will constitute an important sum rule, allowing to quantify the accuracy of our results by assessing how faithfully
the resulting wavefunction is reproduced.  

All the complexity of the problem is therefore hidden in two places.  First, wavefunctions $| \Psi_n\rangle$ and their energy
$E_n$ must be known.  This is standardly handled by the Bethe Ansatz (see references \cite{Bethe1931} - \cite{Takahashi1999}
and references therein;  we provide a summary of the necessary details in Appendices A and B).  
Second, the overlaps $Q_n$ must also be known.  This is a problem of much greater complexity, which in the
present situation finds its solution in the framework of the Algebraic Bethe Ansatz.  The derivation of these overlaps is given
in Appendices C and D.  Note that similar overlaps (quench matrix entries) were also recently calculated using the Algebraic Bethe
Ansatz for the case of the interaction-quenched Richardson model \cite{2009_Faribault_JSTAT_P03018,2009_Faribault_JMP50}.
Here, we also obtain these overlaps using integrability, but using a different method based on the explicit structure of the
solutions to the Bethe equations.

Despite being in possession of these two fundamental building blocks, one major difficulty remains.  Since the Hilbert
space is exponentially large in system size $N$, the summation in (\ref{timeevolphi}) is difficult to handle, and must in
practice be truncated in order to reach sizes sufficiently large to allow the discrimination between finite- and infinite-size
behaviour.  We will show in the next section that this truncation is possible in the situation we consider.  In particular, 
this puts us in position to study the long-time average of observables according to any prescription desired, a common one being 
\begin{equation}
\overline{\langle \mathcal{O} \rangle} \equiv \lim_{T\rightarrow \infty} \frac{1}{T} \int_{0}^T dt \langle \mathcal{O}(t) \rangle,
\end{equation}
and to make reliable observations on the relaxation/thermalization of the prepared state.

The time evolution of this domain wall state has been studied in similar settings. For gapless spin chains this type 
of quench is studied in \cite{Calabrese2008}. An exact analysis has been carried out for the XX-chain in \cite{Antal1999}. 
The short time regime for the XXZ chain is studied using tDMRG in \cite{Gobert2005}, and an exact diagonalization was performed for 
the gapped XXZ chain with open boundary conditions \cite{Haque2009}.  We here offer a complement to these studies, consisting in
results from integrability.

\section{Spectral analysis}
\begin{figure}[h]
\includegraphics[width=7cm]{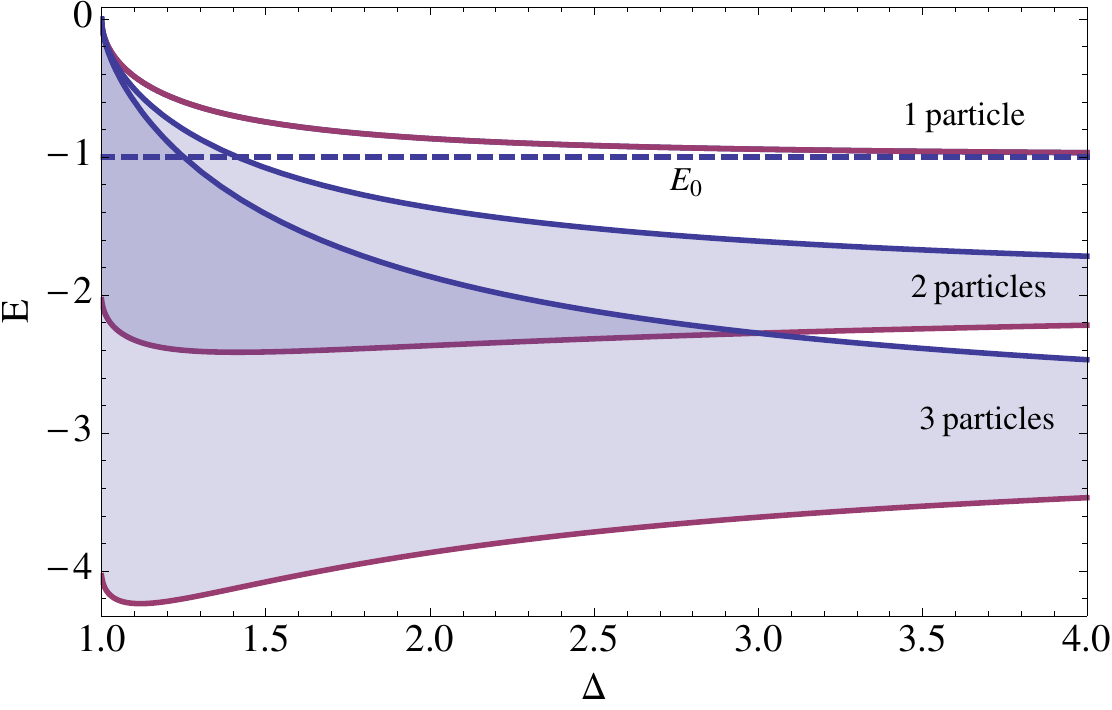}\;
\includegraphics[width=7cm]{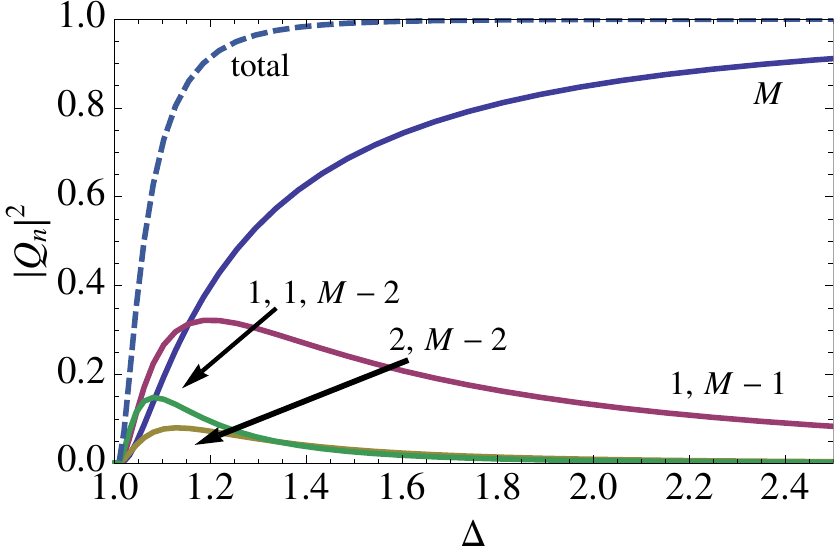}
\caption{Left:  sketch of the energy values covered by the highest three energy bands of the XXZ model as a function of $\Delta$ compared with the energy of the initial state $E_0$. Indicated is the number of particles the states in different bands are made of. Right:
normalization sum rule saturation coming from summing over all states of each basic excitation family, as a function of $\Delta$.  As the gapless regime
is approached, all overlaps scale to zero.}
\label{fig:spectrum_overlaps}
\end{figure}

The space of eigenstates of the XXZ chain is spanned by Bethe wavefunctions, each described by a set of rapidity parameters $\{ \lambda \}$
obtained as solution to the Bethe equations.  Bethe wavefunctions are generically quite complicated objects, and can contain both
unbound and bound magnons.  We refer the reader to the standard literature \cite{Gaudin1983,Korepin1993,Takahashi1999} 
as well as to our appendices for the basic detail.  In summary, solutions of the Bethe ansatz equations can be classified using strings (see Appendix \ref{A:BAE} for a discussion). An $n-$string in a set of $n$ complex rapidities sharing the same real part and invariant under complex conjugation. These strings can be interpreted as bound states of `mass' $n$. In the Ising-limit it can be shown 
that an $n$-string corresponds to $n$ adjacent down spins. 
From a perturbative analysis, we expect at large anisotropy $\Delta$ a well-defined hierarchy of states in terms of the
importance of their overlap with the domain wall state.  This hierarchy is presented in Table \ref{tab:hierarchy}.
The most important states are those consisting of a single $M-$string, which occupy a dispersionless line in the thermodynamic limit.
They are followed by successively 
more complicated partitionings of the $M$ rapidities into more and more individual bound or unbound states, each
partitioning representing a whole continuum of excitations.  The number of particles in the state is defined here to be the number of elements in
the partitioning.
From a perturbative analysis it can be easily shown that for sufficiently large $M$ the system is insensitive to small changes of $M$
({\it i.e.}, of how distant the domain and anti-domain walls are from each other).
This allows us to perform the calculation with an $M$ slightly below $N/2$ without loss of generality. The reason for this choice is that for 
these values the string hypothesis is better satisfied for the majority of string states.  
\begin{table}
\begin{center}
\begin{tabular}{c|c|c|c}
Order $(1/\Delta)$& strings  & \# particles & Energy (Ising limit) \\\hline
0 & $\{M\}$ & $1$ & -1 \\\hline
1 & $\{1,M-1\}$ & $2$ & -2 \\\hline
2 & $\{1_2,M-2\}$ & $3$ & -3 \\
2 & $\{2,M-2\}$ & $2$ & -2 \\\hline
3 & $\{1_3,M-3\}$ & $4$ & -4 \\
3 & $\{1,2,M-3\}$ & $3$ & -3 \\
3 & $\{3,M-3\}$ & $2$ & -2\\\hline
\vdots & & &\\
\end{tabular}
\end{center}
\caption{The hierarchy of states, in order of importance of contributions to the normalization sum rule (\ref{sumrule}).  Each partitioning of $M$ rapidities into strings leads to an independent excitation class, only the few shown here being of relevance to the current setup.}
\label{tab:hierarchy}
\end{table}
In the left panel of figure \ref{fig:spectrum_overlaps}, we give the band structure of the dominant classes of excitations as a function of $\Delta$.  The number of particles corresponds to the number of strings a state is made of. The right panel of the same figure shows the normalization sum rule saturation coming from these excitations.  For sufficiently large values of anisotropy, 
the saturation is essentially complete using these states only.  Approaching the gapless regime however requires increasingly large numbers
of classes of excitations, since all overlaps scale to zero as can be expected from Anderson's orthogonality catastrophe scenario \cite{1967_Anderson_PRL18}. 
A further issue with the gapless limit $\Delta \rightarrow 1$ is that in its vicinity, solutions to Bethe equations involving complex rapidities become
more and more difficult to find, and often require considering deviated strings in detail \cite{Hagemans2007}.  We will not burden ourselves with
these issues here, and only consider systems sufficiently deep in the gapped regime where such deviations are negligible.

\section{Work probability distribution}
\begin{figure}[h]
\includegraphics[width=7cm]{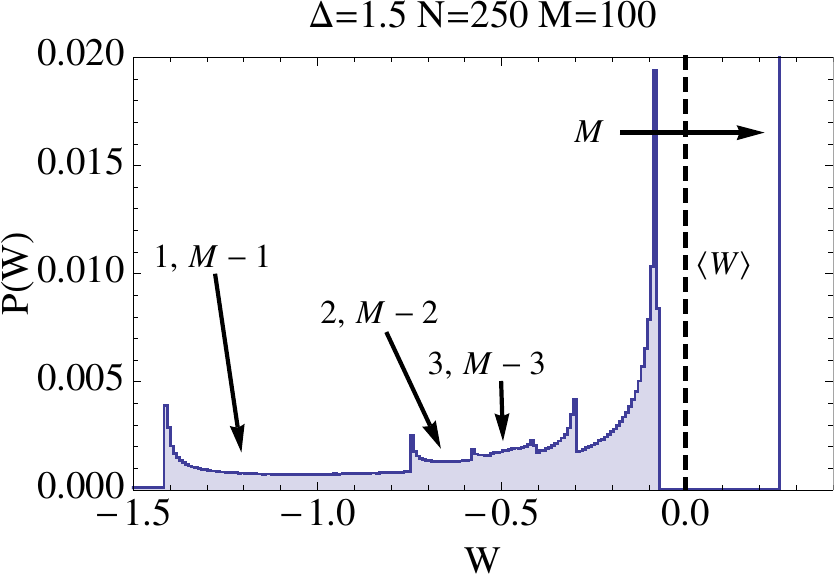}\;
\includegraphics[width=7cm]{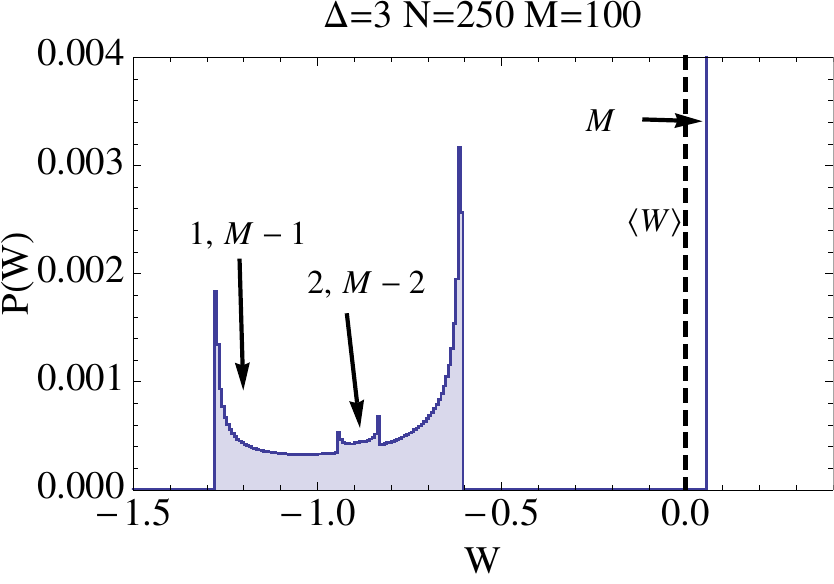}
\caption{Work probability distribution for $N=250$ and $M=100$ and anisotropies  $\Delta=1.5$ (left) and $3$ (right). Indicated are the contributions from the important string states. The vertical scale of the full peak of the $M-$string is cut in the plot for clarity.}
\label{F:workdistr}
\end{figure}

\begin{figure}[h]
\begin{center}
\includegraphics[width=7cm]{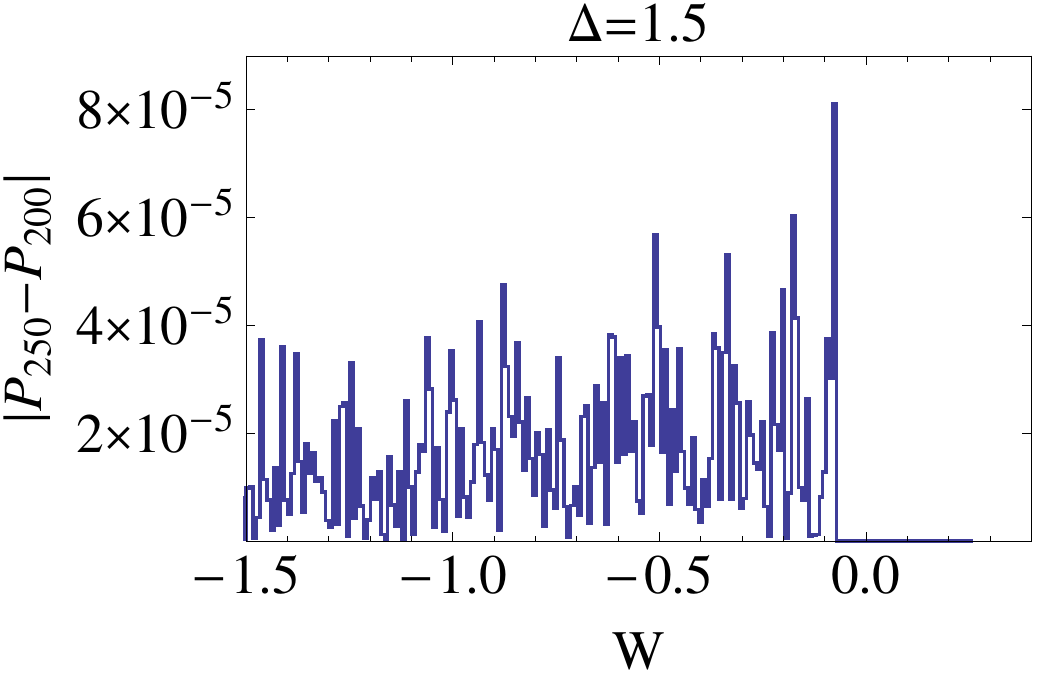}
  \includegraphics[width=7cm]{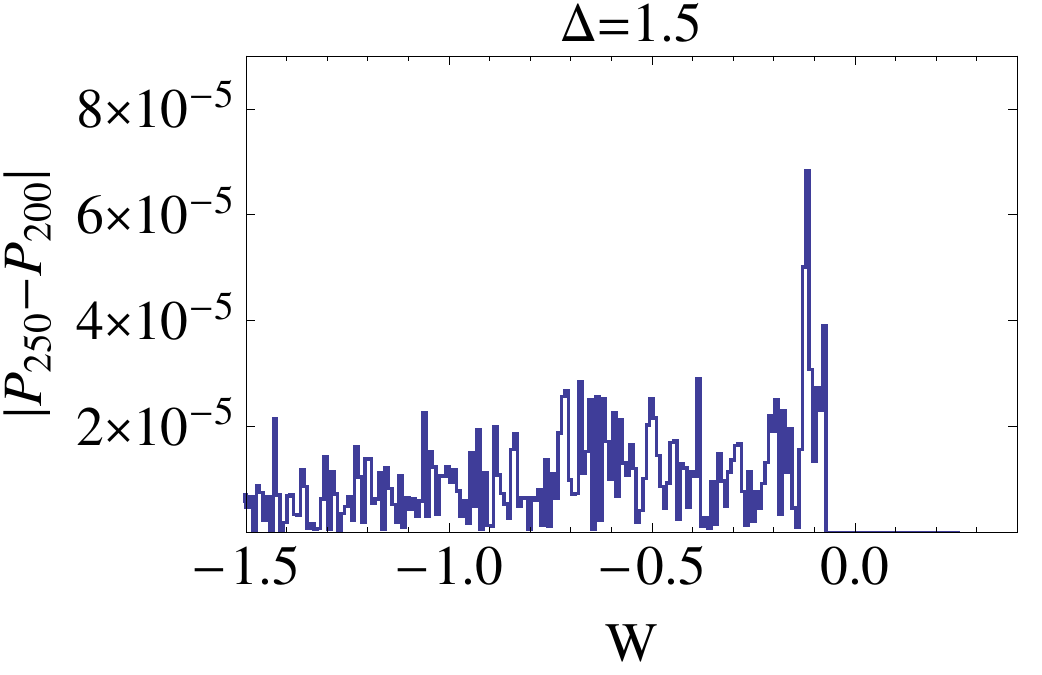}
\caption{For $\Delta=1.5$ we plot the difference of $P(W)$ obtained from different system sizes. 
Left: the difference between $N=200$ and $N=150$. Right: the difference between $N=250$ and $N=200$.
The differences fall off with increasing system size, and for the sizes presented are smaller by about two orders of magnitude 
than the value of $P(W)$ obtained (see figure \ref{F:workdistr}).}
\label{F:workdistrfinitesize}
\end{center}
\end{figure}

One of the most straightforward measurable quantities which can be obtained from the knowledge of the overlaps $Q_n$ 
is the work probability distribution \cite{Silva2008} defined as:
\begin{equation}
P(W) = \sum_{n} |\langle \phi | \Psi_n\rangle |^2 \delta(W-E_n+E_0).
\end{equation}
The work expectation value can easily be computed analytically:  $\langle W \rangle = \langle \phi| H_{XXZ} - H_0| \phi \rangle = 0$ and 
 $\langle W^2 \rangle = \frac{1}{2 \Delta^2}$.  $\langle W \rangle$ can thus be considered intensive here, which for a generic quantum quench is not the case. 
In figure \ref{F:workdistr}
we plot  $P(W)$ for two different values of $\Delta$. The calculations are done for a finite system size ($N=250$, $M=100$), therefore the results are binned in energy
(using bins larger than the interlevel spacing) in order to generate smooth curves.
The contribution of the $M-$strings is a narrow peak with a vanishing bandwidth in the thermodynamic limit, see \eqref{energy}.
The $1,M-1$ states have two peaks at the edge of the energy band. The peak at the right of the continuum close to $\langle W \rangle$ is the result of individual peaks associated to a large 
value of overlap.  The peak at the left of the continuum is the result of smaller overlaps accompanied by an increasing density of states. 
Other states made up of two strings lie within the energy band of the $1,M-1$-strings and display a similar structure.
Since the energies are intensive in this energy domain it is expected that the results for $P(W)$ we obtain closely mimics
the one in the thermodynamic limit. This is motivated by considering the difference in $P(W)$ for two different system sizes, see figure \ref{F:workdistrfinitesize},
showing the fact that finite-size effects rapidly disappear.

\section{Relaxation dynamics: Loschmidt echo}
We now turn to the question of whether the system effectively displays some form of relaxation.  Of course here, since we consider
a single realization and not an ensemble in the presence of a bath, relaxation can only come from the relative dephasing of the
various terms in the right-hand side of (\ref{timeevolphi}).  

\begin{figure}[h]
\includegraphics[height=4.5cm]{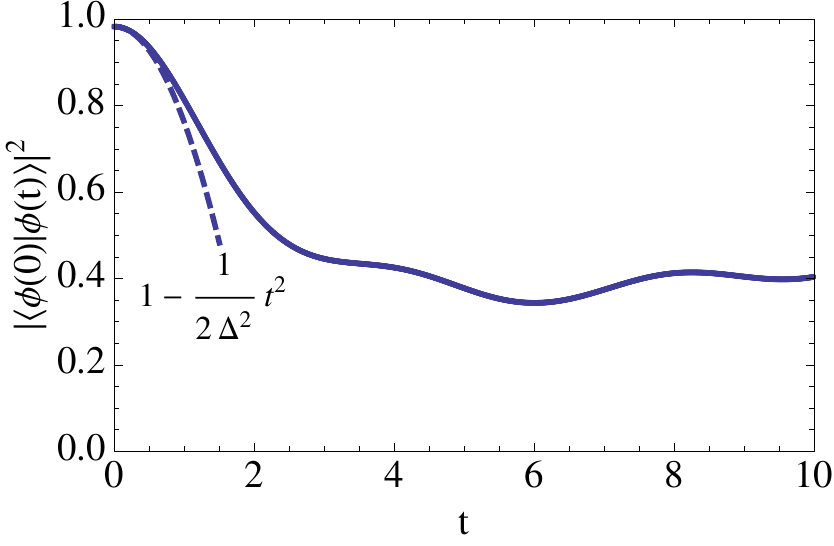}\;
\includegraphics[height=4.5cm]{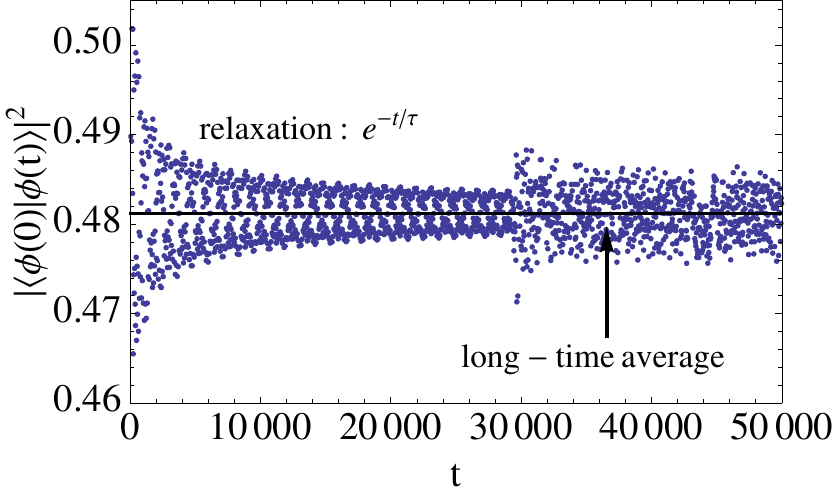}
\caption{Loschmidt echo as a function of time for the domain wall quench for $\Delta=1.5$ and $N=200$, $M=80$.  In the left panel, the short time behaviour is shown.
The dashed line is the result from perturbation theory.  At large times, plotted in the right panel, one can distinguish effective
relaxation towards the long-time average, but also finite-size induced small-scale revivals after a certain point (see main text). 
Note the change of scale in the horizontal axis in going from one figure to the other.}
\label{fig:LE}
\end{figure}

It is not a priori easy to guess the outcome of the time evolution, since the states occupy coherent modes followed by overlapping continua, 
and each state contributes according to its relative overlap.   Moreover, this outcome
can essentially depend on which observable is considered, via the form factor values of the operator concerned, each operator 
favouring specific sets of state combinations.  As a concrete example, we will focus on studying the 
Loschmidt echo \cite{Silva2008, Zanardi2009} which is the overlap between the inital state and the time evolved state,
\begin{align}\nonumber
\mathcal{L}(t) &= \left| \langle \phi | e^{-i H_0t} e^{i Ht} |\phi \rangle \right|^2\\\nonumber
&=\sum_{m,n} e^{i(E_m-E_n)t} |\langle \phi |\Psi_n\rangle|^2 |\langle \phi |\Psi_m\rangle|^2\\
&=\left| \int P(W) e^{i W t} dW  \right|^2.
\end{align}
This will allow us to quantify if the system dephases, if so how quickly and to what point, and whether initial state (partial) 
revival can take place.  Plots of the Loschmidt echo for short and long times are presented in figure \ref{fig:LE}.  
We can separate different regimes in the time dependence.  First and foremost, the small time, transient regime plotted in the
left panel can be easily understood using simple perturbation theory, which predicts an initial decay quadratic in time with an
anisotropy-dependent coefficient:
\begin{align}\nonumber
\mathcal{L}(t) &= 1 + (\langle H \rangle^2 - \langle H^2 \rangle) t^2 + O(t^4)\\
&=1 - \frac{1}{2 \Delta^2}t^2 + O(t^4).
\end{align}
More complex behaviour becomes evident when considering longer times, as in the right panel of figure \ref{fig:LE}.
One point worth emphasizing is that our method, unlike {\it e.g.} real-time numerics, is (in view of the exact
wavefunctions and energies we use) directly applicable irrespective of the time $t$ considered, and thus valid also at large times.
There however, finite-size effects show up, originating from the discrete nature of the energy levels.  Numerical
inaccuracies also prevent us from giving numerical values at arbitrarily large times.  Returning to the figure, a
number of interesting features are displayed.  First, interferences between the various terms lead to 
oscillations around the long-time average
\begin{equation}\label{LTALoschmidt}
\overline{\mathcal{L}} = \sum_{m,n} \delta(E_m-E_n) |\langle \phi |\Psi_n\rangle|^2 |\langle \phi |\Psi_m\rangle|^2.
\end{equation}
These oscillations are accompanied by a decaying envelope which can be interpreted as effective relaxation.
This behaviour carries on until a new regime is entered, where finite-size effects take over and lead to
accidental partial revivals where the Loschmidt echo shows rapid, larger-scale variations.  More generally,
the echo looks chaotic after the onset of these revivals, and the numerical calculations lose their meaningfulness.
The revival time is observed to grow with system size, first approximately linearly with system size, but more
generally in an unspecifiable and nontrivial fashion.

\section{Thermodynamic limit}
The overlaps of the $M-$string states with the initial domain wall state 
can be computed when $N\rightarrow \infty$ both for the case of $M$ finite and for $M\rightarrow\infty$.
When $M\rightarrow\infty$ the $M$-strings become exactly coherent. We can compute a lower bound for  the long time average of the Loschmidt echo $\overline{\mathcal{L}}$. In appendix D we derive:
\begin{equation}
\lim_{N\rightarrow \infty} \sum_{n=1}^N |\langle \phi|\Psi_n^M\rangle|^2 =  \prod_{n=1}^\infty \left( 1 - e^{-2 n \eta} \right)^2,  \qquad \Delta >1,
\end{equation}
in which we have used the definition $\cosh \eta \equiv \Delta$, and
where the summation on the left-hand side is taken over all $M$-string solutions. 
By only taking these contributions into account we obtain a lower bound for \eqref{LTALoschmidt}:
\begin{equation}\label{LBLoschmidt}
\overline{\mathcal{L}} \geq \prod_{n=1}^\infty \left( 1 - e^{-2 n \eta} \right)^4, \qquad \Delta >1.
\end{equation}
This result can be interpreted as the overlap between the initial state and the asymptotic state. The non-zero overlap indicates that the asymptotic state 
keeps most of the spatial anisotropy from the initial state. 
We can conclude that for this type of quench the system will not thermalize in the thermodynamic limit.

In table \ref{finitesizeLoschmidt} we compare this result for $\overline{\mathcal{L}}$ with the one for a finite system with $N=200$. The finite time average is over intervals 
before the finite size effects show up. The finite size result $\overline{\mathcal{L}}_{200}$ exceeds the thermodynamic limit value 
$\overline{\mathcal{L}}_\infty$ by a small value, which can be attributed to the transient regime, which can be neglected in the infinite 
time average.  This clearly suggests that the lower bound \eqref{LBLoschmidt} is the exact value for the infinite time average in the thermodynamic limit. 
\begin{table}
\begin{center}
\begin{tabular}{c|c|c}
$\Delta$ & $\overline{\mathcal{L}}_\infty$ & $\overline{\mathcal{L}}_{200}$\\\hline
1.5 &0.481212 &0.481402\\
2 &0.725941&0.726285\\
3&0.884184&0.884241\\
4&0.936021&0.936077\\
\end{tabular}
\end{center}
\caption{The lower bound for $\overline{\mathcal{L}}$ in the thermodynamic limit compared with long time average for a finite size system ($N=200$ and $M=80$).}
\label{finitesizeLoschmidt}
\end{table}

\section{Discussion and Conclusions}
In this work we studied the relaxation dynamics of the gapped XXZ chain after a quench from an initial domain wall state. The techniques that were used are based on the Algebraic Bethe Ansatz for a finite chain. Some of the results were extended to the thermodynamic limit. The key ingredient for studying this quench was deriving a numerically efficient expression for the overlap between the initial state and eigenstates of the XXZ Hamiltonian. 
Overlaps for different Ising states as initial state could be computed following the same logic, however in the general case of a finite density of domain walls 
this will become an intractable combinatorial problem.

Since the energies of the relevant part of the spectrum are intensive a good comparison with the thermodynamic limit is possible.  One important result is that the bandwidth of the $M-$strings is nonzero for finite systems allowing the long time average of the Loschmidt echo going to zero, something that does not happen in the thermodynamic limit. 
Hence we conclude that the limits $\lim_{T\rightarrow\infty}$ and $\lim_{N\rightarrow\infty}$ do not commute, which was also pointed out in \cite{Zanardi2009}.

In the thermodynamic limit we derived a lower bound for the infinite time average of the Loschmidt echo and argue that this bound is the exact value. This led us to the most important conclusion of this paper, namely that the system for this quench will not thermalize.
The reason why the system does not thermalize can be explained from its spectrum, and does not directly rely on the 
fact that the system is integrable. Imagine for instance a gapped system with lowest (or highest) band having 
zero bandwidth in the thermodynamic limit, such as a system with macroscopically-degenerate ground states.  
Consider a quench such that $\langle W \rangle$ lies right between the lowest 
(or highest) energy band and the next one.  A lower bound for the overlap  $|\langle \phi | \Psi_0 \rangle |^2$ of the lowest band
can then be obtained by simple reasoning.  Let $E_0$ be the energy of the lowest energy band of zero bandwidth. 
We denote the band gap by $E_1-E_0$.  The expectation value of the initial Hamiltonian is $\langle H_0 \rangle = E_0 + \Delta E$. 
We then choose $0 < \Delta E< E_1-E_0$.  In this case we can obtain a lower bound for the total overlap of the $E_0$ states. 
 \begin{align}\nonumber
 E_0 +\Delta E &= \sum_{n} |\langle \phi | \Psi_n \rangle |^2 E_n\\
 &\geq  |\langle \phi | \Psi_0 \rangle |^2 E_0 + (1-|\langle \phi | \Psi_0 \rangle |^2)E_1,
 \end{align}
from which follows the lower bound
 \begin{equation}
 |\langle \phi | \Psi_0 \rangle |^2 \geq 1 - \frac{\Delta E}{E_1-E_0}.
 \end{equation}
In other words, such a quench must occupy the lowest (resp. highest) dispersionless band with $\mbox{O}(1)$ amplitude.
Since these states are dispersionless and separated from other states by a band gap, they cannot dephase, and
therefore cannot effectively relax.  The large-time asymptotic state will therefore always maintain memory of the
ground state occupation amplitude of this band, which is a quench-dependent quantity. Note that such situations trivially
cannot show eigenstate thermalization according to the arguments of \cite{Srednicki1994}, in which such circumstances
were argued to be non-generic and therefore not treated.

The condition used for deriving this lower bound is quite strong. In figure \ref{fig:spectrum_overlaps} one can see that for $1<\Delta<1.5$ that  $\langle W \rangle$ does not lie in the gap anymore, but the overlaps of the $M-$string  are still substantially  big.  
One can also question whether the condition of an isolated peak is essential. From the work probability  distribution it is clear that the contributions of states made up of different strings can be distinguished. Simply looking at energy levels is therefore
insufficient to assess the presence or absence of relaxation, and one must also take into account the values of the wavefunction overlaps.  
If we consider for example  $n$ strings with a diverging length in the thermodynamic limit, they all will become coherent. Although they are 
embedded in a continuum, we can reasonably expect a non-zero height peak of zero width in $P(W)$, which again will prevent thermalization.
 
In the case of evolution under a gapless Hamiltonian, our method loses its efficiency, since all overlaps scale to
zero and the number of states to take into account grows accordingly, preventing an efficient truncation of the sum in \ref{timeevolphi}.
The absence of a gap thus opens the door to dephasing involving arbitrarily complex excitation continua, in 
correspondence with Anderson's orthogonality catastrophe principle.  
Therefore it is expected that $\overline{\mathcal{L}}=0$ in this case, and the possibility of thermalization
remains, although it cannot be quantified here. The presence of a gap is thus determinantal to the long-time asymptotics of dynamics after the quench release of the domain wall state, but is not very sensitive to system size for large enough systems.

To summarize we considered a quench starting from a domain wall state under the time evolution of the gapped XXZ chain. 
We observed that the long time behavior cannot be described by a statistical ensemble. This can be understood by the fact 
that the spectrum is gapped, and does not directly rely on the fact that the system is integrable.  Furthermore we were able 
to assess finite size effects by considering both results for the finite and infinite systems. General conclusions for the 
long time behavior of the XXZ model after quench cannot however be drawn from the results presented here. For instance, the 
two quasi-degenerate ground states of the XXZ chain we are dealing with are separated from all other states by the gap, so one might 
expect a similar effect if one would perform a quench starting from the N\'eel state.  However, the low energies are extensive 
in contrast with the high energy excitations which are intensive.  The effect of the gap is much less pronounced and one might 
at least expect that the staggered magnetization will vanish for all $\Delta >1$, in correspondence with the results of \cite{Barmettler2009}.  
In future work, we will consider this kind of initial state as well as the time evolution of more elaborate observables.

\appendix

\section{Bethe ansatz equations}\label{A:BAE}
In the Algebraic Bethe Ansatz the eigenstates are completely characterized by a set of rapidities $\{\lambda_j\}, j=1\ldots M$, solutions to the Bethe equations.  These solutions can be either real or complex. The complex solutions typically come in groups, invariant under complex conjugation, which represent bound states. At a small density of down spins or large anisotropy $\Delta$, most solutions take the form of so-called strings, in which a number of rapidities share a real centre while the imaginary parts are equally spaced according to the so-called string hypothesis.  An $n-$string takes the form:

\begin{equation}
\lambda^n_{\alpha,j} = \lambda_{\alpha}^n + i\eta (n+1-2j)/2 + i\delta_j \qquad j=1\dots n
\end{equation}
where $\eta = \mbox{acosh}(\Delta)$, and $\delta_j$ is a deviation. When the string hypothesis holds (i.e. when all deviations $\delta_j$ are sufficiently small) the Bethe equations can be rewritten as equations for the real parts of the rapidities only, yielding the Bethe-Gaudin-Takahashi equations \cite{Takahashi1999}. The logarithmic  version of the Bethe-Gaudin-Takahashi equations for $\Delta >1$ is given by
\begin{equation}\label{2:bethetaka}
 \theta_n (\lambda_\alpha^n) = \frac{2\pi}{N} I_\alpha^n + \frac{1}{N}\sum_{(m,\beta)\neq (n,\alpha)} \Theta_{nm} (\lambda_\alpha^n - \lambda^m_\beta)
\end{equation}
with the dispersion and scattering kernels
\begin{equation}\label{2:theta}
 \theta_n (\lambda) = 2\arctan \left( \frac{\tan \lambda}{\tanh n \eta /2}\right) + 2\pi \left\lfloor \frac{\lambda}{\pi}+ \frac{1}{2} \right\rfloor,
\nonumber
\end{equation}
\begin{equation}\label{2:bigtheta}
 \Theta_{mn}(\lambda) \equiv
\begin{cases}
 \theta_{|n-m|}(\lambda) + 2 \theta_{|n-m|+2}(\lambda) + \ldots + 2 \theta_{n+m-2}(\lambda) + \theta_{n+m}(\lambda) &n\neq m\\
2 \theta_2(\lambda) + 2 \theta_4(\lambda) + \ldots + 2 \theta_{2n-2}(\lambda) + \theta_{n+m}(\lambda) &n=m.
\end{cases}
\end{equation}
Solutions of the Bethe-Gaudin-Takahashi equations can be classified by the set of quantum numbers $I_\alpha^n$, which are integers 
(half-odd integers) if $N-M_n$ is odd (even). The partitioning into strings satisfies the constraint 
$\sum_{n=1}^M n M_n = M$. The energy of an $n-$string $\lambda^n_\alpha$ is
\begin{equation}
\epsilon_n(\lambda^n_\alpha) = -  \tanh (\eta)  \frac{\sinh n \eta}{\cosh n \eta - \cos 2 \lambda_\alpha^n}.
\end{equation}
The total energy and momentum of an eigenstate containing strings are given by
\begin{equation}\label{energy}
E = \sum_{\alpha,n} \epsilon_n(\lambda_\alpha^n)
\end{equation}
\begin{equation}
P = \pi \sum_{n} M_n - \frac{2\pi}{N} \sum_{\alpha,n}I^{n}_{\alpha} \quad \mod 2\pi.
\end{equation}
For finite $\eta$ and large $n$ the energy $\epsilon_n(\lambda)$ becomes dispersionless.  An $n-$string can be interpreted as
a magnon bound state with a mass proportional to $n$.

\subsubsection*{Band structure}
In the thermodynamic limit $(N\rightarrow \infty)$ the band structure is easily obtained from \eqref{energy}. 
For a given string configuration $\{M_n\}$ one can define: $E^{max}(\{M_n\})$ by putting all $\lambda^n_\alpha$ to $\pi/2$. Similarly one can define  $E^{min}(\{M_n\})$ by putting all $\lambda^n_\alpha$ to $0$.  The $M-$string is a special case since it has a vanishing bandwidth 
in the limit $M\rightarrow \infty$. In the Ising limit ($\eta \rightarrow \infty$) the energy of a state made of $m$ different strings is $-m$.  
This corresponds in the Ising language to a state made of $m$ distinct blocks of down spins and clarifies the bound states interpretation in this limit.

\section{Classification of string solutions}
In order  to evaluate the spectrum numerically a correspondence between quantum numbers and rapidities is needed. For the isotropic chain $(\Delta =1)$ every set of quantum numbers corresponds to a unique solution of the Bethe-Gaudin-Takahashi equations. A complete classification of solutions in terms of quantum numbers was given in \cite{Takahashi1999}. For the case of the XXZ chain with $\Delta>1$ this classification in not known in general. For solutions with only real solutions this was done in \cite{Caux2008}. Here we generalize the result to string solutions.  Note that a genuine proof that the Bethe-Gaudin-Takahashi equations produce the right number of eigenstates for $\Delta >1$ remains to be found.

\subsubsection*{Bandwidth: $W_n$}
In \cite{Caux2008} it was argued that $\lambda_j - \lambda_k$ if $I_j <I_k$.  Equal rapidities do not yield proper Bethe wavefunctions, and we therefore only need to consider sets of distinct quantum numbers. In the parameterization chosen in the Bethe-Gaudin-Takahashi equations all rapidities are restricted to an interval of width $\pi$, i.e. for all strings of given length $n$ we have:
 $\lambda^n_{M_n}-\lambda^n_1<\pi$. From this requirement and assuming that the Bethe-Gaudin-Takahashi equations are monotonic in $\lambda_\alpha^n$  it follows that: $I^n_{M_n}-I^n_1<N-\sum_m t_{nm} M_m \equiv  W_n$ with $t_{nm} \equiv 2 \mbox{min}(n,m)-\delta_{n,m}$. We call $W_n$ the bandwidth of the quantum numbers $\{I_j^n\}$. 
 
\subsubsection*{Quasi-periodicity of the Bethe-Gaudin-Takahashi equations: $B_n$}
Since $\lambda + \pi \equiv \pi$ as far as the wavefunctions are concerned, there is no one-to-one mapping of quantum numbers and wavefunctions. Namely,  the shift  $\lambda^n_1\rightarrow \lambda^n_1 +\pi$ leaves the wavefunction invariant although the quantum numbers in the Bethe-Gaudin-Takahashi equations change. For every string length $n$ we can define a transformation: $S^n: \{(\lambda_j^a,I_j^a)\} \rightarrow \{(\tilde{\lambda}_j^a,\tilde{I}_j^a)\}$ with:
\begin{align}\nonumber\label{Stransform} 
(\tilde{\lambda}_{M_n}^n, \tilde{I}_{M_n}^n) &= (\lambda_{1}^n\!+\!\pi, I_{1}^n\!+\!(W_n\!+\!(2n\!-\!1)))\\\nonumber
(\tilde{\lambda}_{j}^n,\tilde{I}_j^n) &= (\lambda_{j\!+\!1}^n,I_{j\!+\!1}^n\!+\!(2n\!-\!1)) & j=1 \ldots M_n-1\\
 (\tilde{\lambda}_{j}^m, \tilde{I}_j^m) &= (\lambda_{j}^m,I_{j}^m + t_{nm}) & m\neq n,\; j=1\ldots M_m.
\end{align}
From this transformation we see that $\tilde{I}_1^n \geq I^n_1 + 2n$. For every string length $n$ there are $B_n \equiv 2n$ bands of quantum numbers which cannot be 
transformed into each other by $S^n$, whereas in the XXX 
case there is only one such band for every string length. If we restrict to only one string length it follows that the number of choices for the quantum numbers $\{I^n_j\}$ in a single band is: $\begin{pmatrix}W_n\\M_n\end{pmatrix}$. In case of $B_n$ bands the number of solutions is:
\begin{equation}
\begin{pmatrix}W_n \\M_n\end{pmatrix} + (B_n-1) \begin{pmatrix}W_n-1 \\M_n-1\end{pmatrix}=
\frac{(B_n-1)M_n+W_n}{W_n}\begin{pmatrix}W_n \\M_n\end{pmatrix}.
\end{equation}

\subsubsection*{Degeneracies from different string sectors}
To find the minimal set of quantum numbers one should first determine using $S^n$ the bandwidth $W_n$ and the number of bands $B_n$ for every string length $n$. However, the transformation $S^n$ not only affects the quantum numbers of length $n$ but also all the others. 
To take this effect into account one should also consider transformations of the form $S^m (S^n)^{-1}$, leading to 
additional state exclusions.  In general these are very difficult to determine. The explicit classification for two string types is given below.

\subsubsection*{Two types of strings}
We consider $m$ $n-$strings and $\bar{m}$ $\bar{n}-$strings such that: $m n + \bar{m} \bar{n} = M$. The bandwidths of the two sets of quantum numbers are:
\begin{align}
W_n&= N- (2n-1)m -2n\bar{m}\\
W_{\bar{n}}&=N-2n m-(2\bar{n}-1)\bar{m}.
\end{align}
The number of bands are $B_n = 2n$ and $B_{\bar{n}} = 2\bar{n}$. To take care of the degeneracies coming from different string sectors we consider the transformation:
\begin{equation}
S^{n(\bar{n})^{-1}} \begin{pmatrix} I^n_1 \ldots I^n_m\\I^{\bar{n}}_1 \ldots I^{\bar{n}}_{\bar{m}} \end{pmatrix}=
\begin{pmatrix}I^n_2 -1 \ldots I^n_1 + W_n -1\\I^{\bar{n}}_{\bar{m}} -W_{\bar{n}} -(2\bar{n}-1)+2n \ldots  I^{\bar{n}}_{\bar{m}-1}  -(2\bar{n}-1)+2n \end{pmatrix}.
\end{equation}
This turns out to be the only extra transformation one needs to consider. 
Define: $I^n_{max} = I^n_{min} + (W_n-1)+(B_n-1)$. From the conditions: $I^n_1+W_n-1 \leq I^n_{max}$ and $I^{\bar{n}}_{\bar{m}}   -W_{\bar{n}} -(2\bar{n}-1)+2n \geq I^{\bar{n}}_{min}$ follows that we have to exclude: $I^n_1 \in \{I^n_{min} \ldots I^n_{min} + 2n-1\}$ and $I^{\bar{n}}_{\bar{m}} \in \{I^{\bar{n}}_{max}-2n+1 \ldots I^{\bar{n}}_{max} \}$. The total number of unique solutions for this case is
\begin{multline}
\frac{(B_n-1)m+W_n}{W_n} \begin{pmatrix} W_n\\m\end{pmatrix} 
\frac{(B_{\bar{n}}-1)\bar{m}+W_{\bar{n}}}{W_{\bar{n}}} \begin{pmatrix} W_{\bar{n}}\\\bar{m}\end{pmatrix}-(2n)^2 \begin{pmatrix} W_n-1\\m-1\end{pmatrix}  \begin{pmatrix} W_{\bar{n}}-1\\\bar{m}-1\end{pmatrix}\\
= \frac{N^2-2n(m+\bar{m})N}{W_n W_{\bar{n}}} \begin{pmatrix} W_n\\m\end{pmatrix}   \begin{pmatrix} W_{\bar{n}}\\\bar{m}\end{pmatrix}.
\end{multline}

\section{The overlap matrix}
The Algebraic Bethe Ansatz deals with the problem of diagonalizing simultaneously the transfer matrix $\mathcal{T}(\lambda)$ for all values of $\lambda$, from which all conserved quantities can be obtained \cite{Korepin1993}. All operators can be expressed in terms of four non-local Hilbert space 
operators $A(\lambda), B(\lambda), C(\lambda)$ and $D(\lambda)$. 
The Algebraic Bethe Ansatz requires a pseudo-vacuum $|0\rangle$ which in case of the XXZ chain is the state with all spins up, such that:
\begin{align}\nonumber
A(\lambda) |0\rangle &= a(\lambda) |0\rangle\\\nonumber
D(\lambda) |0\rangle &=d(\lambda) |0\rangle\\\nonumber
B(\lambda) |0\rangle &\neq 0\\
C(\lambda) |0\rangle& =0
\end{align}
The transfer matrix is constructed as: $\mathcal{T}(\lambda) = (A+D)(\lambda)$. The vacuum eigenvalues of the $A(\lambda)$ and $D(\lambda)$ operators are:
\begin{equation}
a(\lambda) =1, \qquad d(\lambda) = \prod_{j=1}^N \frac{\varphi(\lambda - \xi_j)}{\varphi(\lambda -\xi_j +i\eta)},
\end{equation}
\begin{equation}\label{parametrizations}
 \varphi(\lambda+i\eta) = \begin{cases}\lambda+i & \Delta=1\\
 \sinh(\lambda+i\eta) & |\Delta=\mbox{acos}(\eta)|<1\\
 \sin(\lambda+i\eta) & |\Delta=\mbox{acosh}(\eta)|>1.
 \end{cases}\end{equation}
The set $\xi_j$ are  inhomogeneity parameters, which should be set to $\xi_j=i\eta/2$ in order to obtain the XXZ chain.  States, respectively dual states can be constructed as:
\begin{align}\nonumber
|\psi \rangle &=\prod_{j=1}^M B(\lambda_j) |0\rangle\\
\langle \psi | &=\langle 0 | \prod_{j=1}^M C(\lambda_j).
\end{align}
In order to represent an eigenstate the rapidities $\lambda_j$ should satisfy the Bethe equations. These states are not automatically normalized, we therefore write: $|\Psi\rangle = |\psi\rangle/\sqrt{\langle \psi|\psi\rangle}$. The norm of a Bethe state in case of strings is given by \cite{Gaudin1983,Korepin1982,KirillovJMS40,Caux2005}:
\begin{equation}
\langle \psi|\psi\rangle= \varphi(i\eta)^M \prod_{j\neq k,\lambda_j\neq\lambda_k-i\eta} \frac{\varphi(\lambda_j-\lambda_k+i\eta)}{\varphi(\lambda_j-\lambda_k)} \det \Phi^{(r)}(\{\lambda_j\}) + O(\delta)
\end{equation}
where $\Phi^{(r)}$ is called the reduced Gaudin matrix:
\begin{multline}
\Phi_{(j,\alpha)(k,\beta)}^{(r)} = \delta_{jk}\delta_{\alpha\beta} \left(N\frac{d}{d\lambda_\alpha^j} \theta_j(\lambda_\alpha^j) - \sum_{(l,\gamma)\neq (j,\alpha)} \frac{d}{d\lambda_{\alpha}^j} \Theta_{jl}(\lambda_\alpha^j-\lambda_\gamma^l) \right)
+(1-\delta_{jk}\delta_{\alpha\beta}) \frac{d}{d\lambda_\alpha^j}\Theta_{jk}(\lambda_\alpha^j-\lambda_\beta^k).
\end{multline}
An extremely useful formula is Slavnov's expression for the scalar product between an 
eigenstate $\langle \psi (\{\lambda\})|$ (represented by a set of rapidities $\{\lambda\}$ that satisfy the Bethe equations)  and an arbitrary state $|\psi(\{\mu\})\rangle$ (without restrictions on $\{\mu\}$) \cite{Slavnov1989}:
\begin{equation}
\langle \psi(\{\lambda\}) |\psi( \{\mu\} )\rangle = \frac{\det H(\{\lambda_j\},\{\mu_j\})}{\prod_{j>k}\varphi(\lambda_j -\lambda_k) {\prod_{j<k}\varphi(\mu_j -\mu_k)}}
\end{equation}
with
\begin{equation}
H_{ab} = \frac{\varphi(i\eta)}{\varphi(\lambda_a-\mu_b)\varphi(\lambda_a-\mu_b+i\eta)}\left( \prod_{l=1}^M \varphi(\lambda_l -\xi_b+i\eta)  - d(\mu_b)   \prod_{l=1}^M \varphi(\lambda_l -\mu_b-i\eta)\right).
\end{equation}

\subsection{General overlap}
The initial state can be written in the Algebraic Bethe Ansatz language using the inverse mapping \cite{Maillet1999}:
\begin{equation}
S_j^- = \prod_{k=1}^{j-1} (A+D)(\xi_k) B (\xi_j) \prod_{l=j+1}^N (A+D)(\xi_l).
\end{equation}
Making use of $\prod_{j=1}^N (A+D)(\xi_j)  =1$ and $(A+D)(\xi_j)|0\rangle  = |0\rangle$ we can write the state as:
\begin{equation}
|\phi\rangle = \prod_{j=1}^M B(\xi_j) |0\rangle
\end{equation}
which is a normalized state. The overlap between an eigenstate for finite $\Delta$ and the initial state $|\phi\rangle$ can be 
computed using Slavnov's theorem for the scalar product.
In the case when $\{\mu_j\} = \{\xi_j\}$ the matrix $H$ simplifies:
\begin{equation}
H_{ab} = \frac{\varphi(i\eta)}{\varphi(\lambda_a-\xi_b)\varphi(\lambda_a-\xi_b+i\eta)} \prod_{l=1}^M \varphi(\lambda_l -\xi_b+i\eta).
\end{equation}
For the XXZ spin chain all inhomogeneity parameters should be set to $\xi_j = i\eta/2 \; \forall \; j$. The limit should be taken carefully,
there are $M(M-1)$ poles coming from:  $\prod_{j<k}\varphi(\xi_j -\xi_k)$.
Since the columns of the  matrix $H$  are also $M(M-1)$-fold degenerate we can apply l'H\^{o}pital's rule. First we notice that we can easily extract factors $\prod_{l=1}^M \varphi(\lambda_l -\xi_b+i\eta)$ from the determinant, since applying l'H\^opital's rule to those terms does not remove any degeneracies.
\begin{equation}
\langle \psi | \phi \rangle =  \prod_{k=1}^M \prod_{l=1}^M \varphi(\lambda_l -\xi_k+i\eta) \frac{\det \tilde{H}(\{\lambda_j\},\{\xi_j\})}{\prod_{j>k}\varphi(\lambda_j -\lambda_k) {\prod_{j<k}\varphi(\xi_j -\xi_k)}},
\end{equation}
\begin{equation}
\tilde{H}_{ab} = \frac{\varphi(i\eta)}{\varphi(\lambda_a-\xi_b)\varphi(\lambda_a-\xi_b+i\eta)}.\end{equation}
Note that this determinant of $\tilde{H}$ is the same as that in the partition function of the 6-vertex model 
with domain wall boundary conditions \cite{Izergin1987}. However in this case both the sets $\{\lambda_j\}$ and $\{\xi_j\}$ 
play the role of boundary conditions and satisfy no Bethe equations.
To be able to apply  l'H\^opital we are interested in $\partial_{\xi_b}^n \tilde{H}_{ab}$. First we rewrite $\tilde{H}_{ab}$:
\begin{equation}
\tilde{H}_{ab} = \frac{\varphi'(\lambda_a-\xi_b)}{\varphi(\lambda_a-\xi_b)}-\frac{\varphi'(\lambda_a-\xi_b+i\eta)}{\varphi(\lambda_a-\xi_b+i\eta)}.
\end{equation}
Next, consider
\begin{equation}\label{diffH}
\partial_{\xi_b} \left(\frac{\varphi'(\lambda_a-\xi_b)}{\varphi(\lambda_a-\xi_b)}\right)^n =  n  \left(\frac{\varphi'(\lambda_a-\xi_b)}{\varphi(\lambda_a-\xi_b)}\right)^{n+1} - n  \left(\frac{\varphi'(\lambda_a-\xi_b)}{\varphi(\lambda_a-\xi_b)}\right)^{n-1}.
\end{equation}
Using this result it is easy to derive for $n$ even:
\begin{equation}\label{nth-diff}
\partial_{\xi_b}^n\tilde{H}_{ab} =    \sum_{j=0}^{n/2} c_j^n  \left\{ \left(\frac{\varphi'(\lambda_a-\xi_b)}{\varphi(\lambda_a-\xi_b)}\right)^{2j+1}  -  \left(\frac{\varphi'(\lambda_a-\xi_b+i\eta)}{\varphi(\lambda_a-\xi_b+i\eta)}\right)^{2j+1}  \right\},
\end{equation}
and for $n$ odd we get a similar result.
In general the coefficients $c^n_j$ are complicated expressions. However it is straightforward to derive that $c_{n/2}^n = n!$ for $n$ even. 
Since in this case there is no mixing between the left and right terms in \eqref{diffH}. It will turn out that theses coefficients are the only ones we need, 
since all terms   with coefficient $c_{j}^n$ for $j<n/2$ can be removed using row manipulations while leaving the determinant invariant.  Differentiating the denominator: $\prod_{j<k}\varphi(\xi_j -\xi_k)$ is much simpler. First we put $\xi_1 = i\eta/2$. If we now consider $\xi_2$ we get a single zero  in the limit $\xi_2 = i\eta/2$. We can continue this logic and see that $\xi_j$ gives rises to $j\!-\!1$ zero's, hence we get a factor $(j\!-\!1)!$. 
We also notice that the factors $n!$ from the numerator cancel against the factors from the denominator so we are left with:
\begin{align}\label{overlap}
\langle \psi | \phi \rangle &=   \prod_{l=1}^M \varphi(\lambda_l +i\eta/2)^M \frac{\det \bar{H}}{\prod_{j>k}\varphi(\lambda_j -\lambda_k) }\\
\bar{H}_{ab} &=   \left(\frac{\varphi'(\lambda_a-i\eta/2)}{\varphi(\lambda_a-i\eta/2)} \right)^b - \left(\frac{\varphi'(\lambda_a+i\eta/2)}{\varphi(\lambda_a+i\eta/2)} \right)^b.
\end{align}
The matrix $\bar{H}$ has a similar structure to a Vandermonde determinant. This makes the matrix ill-conditioned and is 
therefore not suitable for a numerical evaluation. However, for a given string structure the determinant can be written 
in terms of sums of Vandermonde determinants. In the following sections this is done explicitly for the most important cases.
The normalized overlap is:

\begin{align}\label{noverlap}
\frac{\langle \psi | \phi\rangle}{\sqrt{\langle \psi | \psi \rangle}} = \frac{  \prod_{l=1}^M \varphi(\lambda_l +i\eta/2)^M \det \bar{H}}{ \sqrt{-1^{M(M-1)/2}  \varphi(i\eta)^{M}\prod_{j\neq k}\varphi(\lambda_j-\lambda_k+i\eta) \det \Phi}}.
\end{align}
These expressions hold for all three parameterizations in \eqref{parametrizations}. 

\subsection{Overlap in case of an $M-$string}
In the case where the solution is an $M-$string: $\lambda_{\alpha,j} = \lambda_\alpha + i\eta (M+1-2j)/2$ the determinant of 
$\bar{H}$ is exactly  a  Vandermonde determinant. A Vandermonde matrix is defined as: $V_{jk} = x_j^{k-1}$ with all $x_j$ distinct. Its determinant is given by:
\begin{equation}
\det V  = \prod_{j<k} (x_j-x_k).
\end{equation}
Introducing the notation
\begin{equation}
a_j \equiv \frac{\varphi'(\lambda_j+i\eta/2)}{\varphi(\lambda_j+i\eta/2)}, \quad j=1\ldots M+1,
\end{equation}
the determinant of $\bar{H}$ can be written as an  $M+1 \times M+1$ Vandermonde matrix determinant:
\begin{align}\label{det_Mstring}\nonumber
\det \bar{H}&=
\det \begin{pmatrix}
a_2-a_1 & \hdots & a_2^M-a_1^M\\
\vdots & \ddots& \vdots\\
a_{M+1}-a_{M} & \hdots & a_{M+1}^M-a_{M}^M\\
\end{pmatrix}= \det
\begin{pmatrix}
1&a_1 & \hdots & a_1^M\\
\vdots&\vdots & \ddots& \vdots\\
1&a_{M+1} & \hdots & a_{M+1}^M \nonumber
\end{pmatrix}\\\nonumber
&=\prod_{j<k}^{M+1} \left(\frac{\varphi'(\lambda_\alpha^M + i\eta (M+2-2j)/2)}{\varphi(\lambda_\alpha^M + i\eta (M+2-2j)/2)}-\frac{\varphi'(\lambda_\alpha^M + i\eta (M+2-2k)/2)}{\varphi(\lambda_\alpha^M + i\eta (M+2-2k)/2)}  \right)\\
&=\frac{\prod_{n=1}^M  \varphi(ni\eta)^{M+1-n}}{\prod_{j=1}^{M+1}\varphi(\lambda_{\alpha,j}^M + i\eta/2)^M}.
\end{align}
The reduced Gaudin determinant for an M-string takes the simple form:
\begin{equation}
\det \Phi^{(r)}_M = N \frac{\varphi(i \eta M)}{\varphi(\lambda_{\alpha}^M - iM\eta/2) \varphi(\lambda_{\alpha}^M+iM\eta/2)}.
\end{equation}
This results in the normalized overlap for an M-string:
\begin{equation}
\frac{\langle \psi | \phi \rangle}{\sqrt{\langle \psi | \psi \rangle}} = \frac{\prod_{n=1}^{M-1} \varphi(i n \eta)}{\varphi(\lambda_\alpha - i\eta M/2)^M} \sqrt{\frac{\varphi(\lambda_\alpha -i\eta M/2)\varphi(\lambda_\alpha+i\eta M/2)}{-1^{M(M-1)/2 }N}}
\end{equation}
The absolute value squared of the overlap is
\begin{align}
\frac{|\langle \psi | \phi \rangle|^2}{\langle \psi |\psi \rangle} &=   \frac{\prod_{n=1}^{M-1} |\varphi(in\eta)|^2}{N(\varphi(\lambda_\alpha-i \eta M/2)\varphi(\lambda_\alpha+i \eta M/2))^{M-1}}.
\end{align}

\subsection{Overlap for two $(m,\bar{m})$-strings}
We consider states constructed from two strings with lengths $m$ and $\bar{m}$ such that $m+\bar{m}=M$, and $m\leq \bar{m}$:
$\lambda^m_j=\lambda^m+i\eta(m+1-2j)/2$ and $\lambda_k^{\bar{m}}=\lambda^{\bar{m}}+i\eta(\bar{m}+1-2k)/2$.  Some convenient notations are:
\begin{align}
a_j &\equiv \frac{\varphi'(\lambda^m_j+i\eta/2)}{\varphi(\lambda^m_j+i\eta/2)}  \qquad j=1\ldots m+1\\
b_j &\equiv \frac{\varphi'(\lambda^{\bar{m}}_j+i\eta/2)}{\varphi(\lambda^{\bar{m}}_j+i\eta/2)}\qquad j=1\ldots \bar{m}+1.
\end{align}
Now the overlap matrix takes the form:
\begin{equation}
\det  \begin{pmatrix}
a_2-a_1 & \hdots & a_1^M-a_1^M\\
\vdots & \ddots& \vdots\\
a_{m+1}-a_{m} & \hdots & a_{m+1}^M-a_{m}^M\\
b_2-b_1 & \hdots & b_2^M-b_1^M\\
\vdots & \ddots& \vdots\\
b_{\bar{m}+1}-b_{\bar{m}} & \hdots & b_{\bar{m}+1}^M-b_{\bar{m}}^M\\
\end{pmatrix}
=\det \begin{pmatrix}
1 & 1 & a_1 & \hdots & a_1^M\\
\vdots & \vdots & \vdots &\ddots &\vdots\\
1 & 1 & a_{m+1} & \hdots& a_{m+1}^M\\
0 & 1 & b_1 & \hdots &b_1^M\\
\vdots & \vdots & \vdots &\ddots &\vdots\\
0 & 1 & b_{\bar{m}+1} & \hdots &b_{\bar{m}+1}^M\\
\end{pmatrix}
\end{equation}
Expanding this determinant in the first column  results in a sum over   $m+1$ Vandermonde determinants. After factoring out common factors we obtain:
\begin{equation}
\det \bar{H} =\sum_{j=1}^{m+1} \left( \prod_{k\neq j}^{m+1} \prod_{l=1}^{\bar{m}+1} (a_k-b_l)  \prod_{\tiny \begin{array}{c}p<q  \\p,q\neq j\end{array}}^{m+1} (a_p-a_q) \right) \prod_{r<s}^{\bar{m}+1}(b_r-b_s).
\end{equation}

\subsection{Overlap for a general string solution}
Consider a state constructed of $n$ strings with lengths: $\l_1 \leq l_2 \ldots \leq l_n$.  Introducing the notation
\begin{align}
a_j^\alpha &\equiv \frac{\varphi'(\lambda^{l_\alpha}_{\alpha,j}+i\eta/2)}{\varphi(\lambda^{l_\alpha}_{\alpha,j}+i\eta/2)}\qquad j=1\ldots l_\alpha +1 \quad \alpha=1\ldots n,
\end{align}
the overlaps can be written as a sum over Vandermonde determinants:
\begin{equation}
\det \bar{H} =\sum_{j_1=1}^{l_1+1} \ldots \sum_{j_{n-1}=1}^{l_{n-1}+1} \left( \prod_{\alpha < \beta}^n \prod_{j\neq j_\alpha}^{l_\alpha+1} \prod_{k\neq j_\beta}^{l_\beta+1} (a_j^\alpha-a_k^\beta)
\prod_{\gamma=1}^{n} \prod_{j<k\; j,k\neq j_\gamma}^{l_\gamma+1} (a_j^\gamma-a_k^\gamma)\right).
\end{equation}

\section{Thermodynamic limit}
The Bethe-Gaudin-Takahashi equations for the $M-$strings reduces to the simple equation:
\begin{equation}
\theta_M (\lambda_\alpha^M)  = \frac{2\pi}{N} I_{\alpha}^M
\end{equation}
For even $M$ and $\Delta>1$ the $N$ possible quantum numbers $I_\alpha^M$ are $ \{-(N-1)/2,-(N-1)/2+1,\ldots,(N-1)/2\}$ which follows from \eqref{Stransform}.
In the thermodynamic limit  we can define a density function $\rho(\lambda)$ for an M-string solution centered around $\lambda$:
\begin{equation}
\rho(\lambda^M) = \frac{N}{2\pi} \frac{d}{d\lambda^M} \theta_M(\lambda^M) = \frac{N}{2\pi} \frac{-i\varphi(i \eta M)}{\varphi(\lambda^M-i\eta M/2)\varphi(\lambda^M+i\eta M/2)}
\end{equation}
such that $\int_{-\pi/2}^{\pi/2} \rho(\lambda)  d\lambda = N$. This function $ \rho(\lambda)$ should be interpreted as the number of solutions there are with a rapidity in the domain $\lambda, \lambda + d\lambda$. The total contribution of all $N$ $M-$string solutions is:
\begin{align}\nonumber
\lim_{N\rightarrow\infty} \sum_{l=1}^N \frac{|\langle \psi_l^M|\phi\rangle |^2}{\langle \psi_l^M|\psi_l^M\rangle}  &= \int_{-\pi/2}^{\pi/2}  \frac{\prod_{n=1}^{M-1} |\varphi(in\eta)|^2}{N(\varphi(\lambda-i \eta M/2)\varphi(\lambda+i \eta M/2))^{M-1}} \rho(\lambda)d\lambda \\
&= \frac{-i\varphi(i\eta M)\prod_{n=1}^{M-1} |\varphi(in\eta)|^2}{2 \varphi'(i M \eta/2)^{2M}} \:_2F_1(\frac{1}{2},M,1,\varphi'(i\eta M/2)^{-2})
\end{align}
where $\:_2F_1(a,b,c,z)$ is the hypergeometric function:
\begin{equation}
\;_2F_1(a,b,c,z) = \sum_{n=0}^{\infty} \frac{(a)_n (b)_n}{(c)_n}\frac{z^n}{n!} \qquad (x)_n = x(x+1)(x+2)\ldots(x+n-1).
\end{equation}
In the limit when both $M$ and $N$ are sent to infinity a simpler expression can be obtained:
\begin{align}\nonumber
\lim_{M,N\rightarrow \infty} \sum_{l=1}^N
\frac{|\langle \psi_l^M |\phi\rangle |^2}{\langle \psi_l^M|\psi_l^M \rangle} &=\lim_{M\rightarrow \infty}
\int_{-\pi/2}^{\pi/2}   \frac{\prod_{n=1}^{M-1} |\varphi(in\eta)|^2}{(\varphi(i \eta M)/2)^{M-1}} \frac{-i\varphi(i\eta M)}{\pi\varphi'(i\eta M)} d\lambda+\mathcal{O}\left(\frac{1}{\varphi'(i\eta M)} \right)\\\nonumber
&=\lim_{M\rightarrow \infty}\prod_{n=1}^{M-1}  \frac{|\varphi(in\eta)|^2}{\varphi'(i \eta M)/2}+\mathcal{O}\left(\frac{1}{\varphi'(i\eta M)} \right) \\
&=\prod_{n=1}^{\infty} \left(1- e^{-2n \eta} \right)^2.
\end{align}

\subsubsection*{Thermodynamic limit: XXX}
The isotropic limit $\Delta \rightarrow 1$ should be taken with care, since some M-strings solutions correspond to different string configurations.
For the XXX case we have the following restriction for the quantum number of an M-string: $|I^M|\leq (N-M)/2$. From the Bethe-Gaudin-Takahashi equation:
\begin{equation}
\theta_M(\lambda^M) = 2\pi I^M/N
\end{equation}
we see that in the limit $N\rightarrow \infty$ and finite $M$ allowed values for $\lambda$ go from minus to plus infinity. In this limit we can derive the weight function:
\begin{equation}
\rho(\lambda^M) =\frac{N}{2\pi} \frac{d}{d\lambda^M} \theta_M(\lambda^M) = \frac{N}{2\pi} \frac{M}{(\lambda^M-iM/2)(\lambda^M+iM/2)},
\end{equation}
so the contribution of the $M$-string for finite $M$ becomes:
\begin{align}\nonumber
\lim_{N\rightarrow \infty} \sum_{l=1}^{N-2M} \frac{|\langle \psi_l^M|\phi\rangle |^2}{\langle \psi_l^M|\psi_l^M\rangle}  &=\int_{-\infty}^{\infty} d\lambda \frac{(M-1)!^2}{N((\lambda-iM/2)(\lambda+iM/2))^{M-1}}\rho(\lambda)\\
&= \frac{(2M-2)!}{M^{2M-2}}
\end{align}

\section*{Acknowledgements}
Both authors gratefully acknowledge support from the Stichting voor Fundamenteel Onderzoek der Materie (FOM) in the Netherlands. 
We thank G. Palacios for useful discussions.

\end{document}